\documentstyle[multicol,aps,epsfig,floats]{revtex}

\begin{document}

\twocolumn[\hsize\textwidth\columnwidth\hsize\csname@twocolumnfalse\endcsname

\title{The Droplet State and the Compressibility Anomaly in
  Dilute 2D Electron Systems}
\author{Junren Shi and X. C. Xie}
\address{ Department of Physics, Oklahoma State University,
  Stillwater, OK 74078 } 
\maketitle
\begin{abstract}
  We investigate the space distribution of carrier density and the
  compressibility of two-dimensional (2D) electron systems by using the local
  density approximation. The strong correlation is
  simulated by the local exchange and correlation energies. A slowly
  varied disorder potential is applied to simulate the disorder
  effect. We show that the compressibility anomaly observed in 2D
  systems which accompanies the metal-insulator transition can be
  attributed to the formation of the droplet state due to
  disorder effect at low carrier densities.
\end{abstract}
\bigskip]
The recent discovery\cite{Sergey} of a two-dimensional (2D) metal-insulator
transition (MIT) has raised the important question concerning the
existence of a metallic phase in 2D systems.
In contrast to the scaling theory of localization\cite{scal},
which predicts that only an insulating phase exists in 2D,
there is strong experimental evidence\cite{abrahams} 
for metallic-like behavior in
many 2D samples. This should not be totally surprising
because the dominant Coulomb interaction in these systems may
invalidate the non-interacting scaling theory.
These intriguing experiments generate renewed interests in studying
the properties of low-density 2D electron systems, especially in
the combined effects of interaction and disorder in such 
systems\cite{abrahams}. Most experimental work in the past has concentrated
on transport measurements. Some recent experimental 
studies\cite{Jiang,Ilani} on thermodynamic properties, such as
compressibility $\kappa$ in 2D systems, have shed further light on
understanding the 2D MIT. It is found\cite{Jiang} that 
the negative $1/\kappa$ at low densities
reaches a minimum value at a certain density $n$, and then
increases dramatically with further decreasing $n$.
Although this 
surprising upturn of $1/\kappa$ (compressibility anomaly)
was observed much earlier in a pioneering
work by Eisenstein {\it et al.}\cite{Eisenstein}, this is the
first time that the minimum point in $1/\kappa$ is identified as
the critical density for the 2D MIT\cite{Jiang}. On the theory side,
there are recent efforts\cite{Si,Pastor} in addressing
the interplay between interaction and disorder, and their effect
in thermodynamic properties. 

In this paper we investigate the space distribution of carrier density and the
compressibility of 2D electron systems by using the local
density approximation. The strong correlation in such systems is
simulated by the local exchange and correlation energies. A slowly
varied disorder potential is applied to simulate the disorder
effect. We find that at low average densities electrons form a droplet state 
which is a co-existence phase
of high and low density regions. 
We show that the compressibility anomaly observed in 2D
systems that accompanies the metal-insulator transition can be
attributed to the formation of the droplet state\cite{Shi}.

To investigate the density distribution of a disordered 2D electron
system, we use the density functional theory. The total energy
functional reads
$$
E[{\bf n}]=E_{T}[{\bf n}]+E_{ee}[{\bf n}]+E_{d}[{\bf n}]
          +E_{x}[{\bf n}]+E_{c}[{\bf n}].
$$
Here $ E_{T}[{\bf n}] $ is the functional of the kinetic energy, 
$E_{ee}[{\bf n}]$ is the direct
Coulomb energy due to the charge inhomogeneity and
$E_{d}({\bf n})$ is the potential energy due to the disorder. The strong
correlation effect caused by the electron-electron interaction is
included in the final two terms: $E_{x}[{\bf n}]$ is the exchange energy
and $E_{c}[{\bf n}]$ is the correlation energy. The ground state density
distribution can be obtained by minimizing the total energy functional
with respect to the density.

Using the local density approximation,
the total exchange and correlation energies are written as
$$
E_{x(c)}[{\bf n}]\approx \int d{\bf x}
\epsilon^{0}_{x(c)}\left[ {\bf n}({\bf x}) \right]
{\bf n}({\bf x}),
$$ 
where $\epsilon _{x(c)}^{0}(n)$ is the exchange (correlation) energy
density for a homogeneous 2D electron system at a given density $n$, which
can be determined by quantum Monte-Carlo calculations.
In this paper, we use the result from Tanatar
and Ceperley \cite{Tanatar},
\begin{eqnarray*}
\epsilon _{x}^{0}(n) & = & -\frac{8}{3}\sqrt{\frac{2}{\pi}}\sqrt{n}\\
\epsilon _{c}^{0}(n) & = & a_{0}\frac{1+a_{1}x}
{1+a_{1}x+a_{2}x^{2}+a_{3}x^{3}}
\end{eqnarray*}
where $ x=1/(\pi n)^{1/4} $. The energy unit is $1Ry^*=m^*e^{4}/2
\varepsilon^{*2}\hbar ^{2} $, and the parameters $a_{0}=-0.3568 $, $
a_{1}=1.13 $, $ a_{2}=0.9052 $, $a_{3}=0.4165$.

The kinetic energy functional can be written as
$$
E_{T}[{\bf n}]=\int d{\bf x}\sum _{i}\psi ^{\dagger
  }_{i}({\bf x})\left( -\nabla ^{2}\right) \psi _{i}({\bf x}),
$$
where the sum is over all occupied quasi-particle energy levels, and ${\bf
  n}({\bf x})=\sum _{i}|\psi _{i}({\bf x})|^{2} $. 
To further simplify the calculation, we make an approximation to the
kinetic energy so that it can be written in the form of a density
functional \cite{Wang},
$$
E_{T}[{\bf n}] \approx \int d{\bf x}\left[ \pi {\bf n}({\bf x})^{2}
+\frac{1}{4}\frac{|\nabla {\bf n}({\bf x})|^{2}}
{{\bf n}({\bf x})}+\cdots \right]. 
$$
The first term provides the local density approximation for the
kinetic energy, while the second term includes the effect of the
density gradient. The approximation provides enough accuracy for this
class of calculations.

The energy functional for the disorder potential $V_{d}({\bf x})$
can be written as
$$
E_{d}[{\bf n}]=\int d{\bf x}V_{d}({\bf x}){\bf n}({\bf x})\, .
$$
In a real
system, a disorder potential may be slowly varying and has
correlation between different positions. To simulate the situation,
we assume the correlation for the disorder follows the simple
behavior,
$$
\left\langle V_{d}({\bf x})V_{d}({\bf x}')\right\rangle =
V_{s}^{2}\exp \left( -\frac{|{\bf x}-{\bf x}'|}{\xi }\right) \, ,
$$
where $V_{s}$ is the amplitude of the potential fluctuation, and
$\xi$ is the correlation length of the disorder. $\xi$
is roughly the average size of valleys in a disorder landscape.

In summary, the total energy functional is of the form 
\begin{eqnarray*}
E[{\bf n}]=& & \int d{\bf x}\left[ \pi {\bf n}({\bf x})^{2}
   +\frac{1}{4}\frac{|\nabla {\bf n}({\bf x})|^{2}}{{\bf n}({\bf x})}
   +\int d{\bf x}'\frac{n({\bf x}){\bf n}({\bf x}')}
  {|{\bf x}-{\bf x}'|} \right. \\
   & & \left. +V_{d}({\bf x}){\bf n}({\bf x})
  +\epsilon_{x}^{0}({\bf n}){\bf n}({\bf x})+
\epsilon_{c}^{0}({\bf n}){\bf n}({\bf x}) \right] \, .
\end{eqnarray*}
Thus, the
local density approximation converts the strong-interacting problem to
a single particle problem with a self-consistently determined
potential. 
The density distribution of the ground state can be obtained by
minimizing the energy functional under the constraint of a constant
total electron number.  We introduce the variable $\chi$ with 
${\bf n}({\bf x})\equiv N\chi ({\bf x})^{2}/\int d {\bf x}'\chi
({\bf x}')^{2} $, where $ N $ is the total number of the
electrons in the system. The constraint for the constant total
electron number is automatically satisfied with the new variable.  We
can get the minimized energy functional by using steepest descent
method with iterations \cite{Wang},
$$
\chi ^{m+1}({\bf x})=\chi ^{m}({\bf x})-\gamma 
\frac{\delta E[\chi ]}{\delta \chi }\left| 
_{\chi =\chi ^{m}({\bf x})}\right.
\, ,
$$ 
where $ \gamma $ is the iteration constant which is chosen so that
the interaction is convergent, and
$$
\frac{\delta E[\chi ]}{\delta \chi }=\frac{2N}{\int d{\bf x}\chi
  ({\bf x})^{2}}\left[ -\nabla ^{2}+V_{eff}[n]-E_{0}\right] \chi
({\bf x}) ,
$$ 
where
$$
V_{eff}[{\bf n}] =  2\pi n+V_{d}({\bf x})
+\frac{\delta }{\delta {\bf n}}
\left[ \epsilon _{x}^{0}[{\bf n}]{\bf n}+\epsilon _{c}^{0}[{\bf n}]{\bf n}
\right], 
$$
and
$$
E_{0}  =  \frac{\int d{\bf x}\chi ({\bf x})
\left[ -\nabla ^{2}+V_{eff}[{\bf n}]\right] \chi ({{\bf x}})}
{\int d{\bf x}\chi ({\bf x})^{2}}.
$$ 

The calculation is carried out in a $ 128\times 128 $ discrete space.
The size of the system is set as $ L=256a_{B}^* $, where $a_{B}^*$ is the
effective Bohr's radius, $ a_{B}^*=\varepsilon \hbar
^{2}/m e^{2} $, with $\varepsilon$ being the dielectric constant
and $m$ the effective mass of an electron.  The periodic
boundary condition and the Ewald sum for the Coulomb interaction are
applied to minimize the finite size effect. The electron density is
adjusted by changing the total electron number $ N $. The density
distribution and the total energy of the system are calculated for
different densities, and the chemical potential is calculated by 
using the formula,
$$
\mu (N)=E(N+1)-E(N)\, .
$$
The compressibility of the system can be calculated by 
$$
\frac{1}{\kappa }=\frac{N^{2}}{S}\frac{\partial \mu }{\partial N}\, ,
$$
where $ S $ is the total area of the system. 

\begin{figure}
  \centering
  \epsfig{file=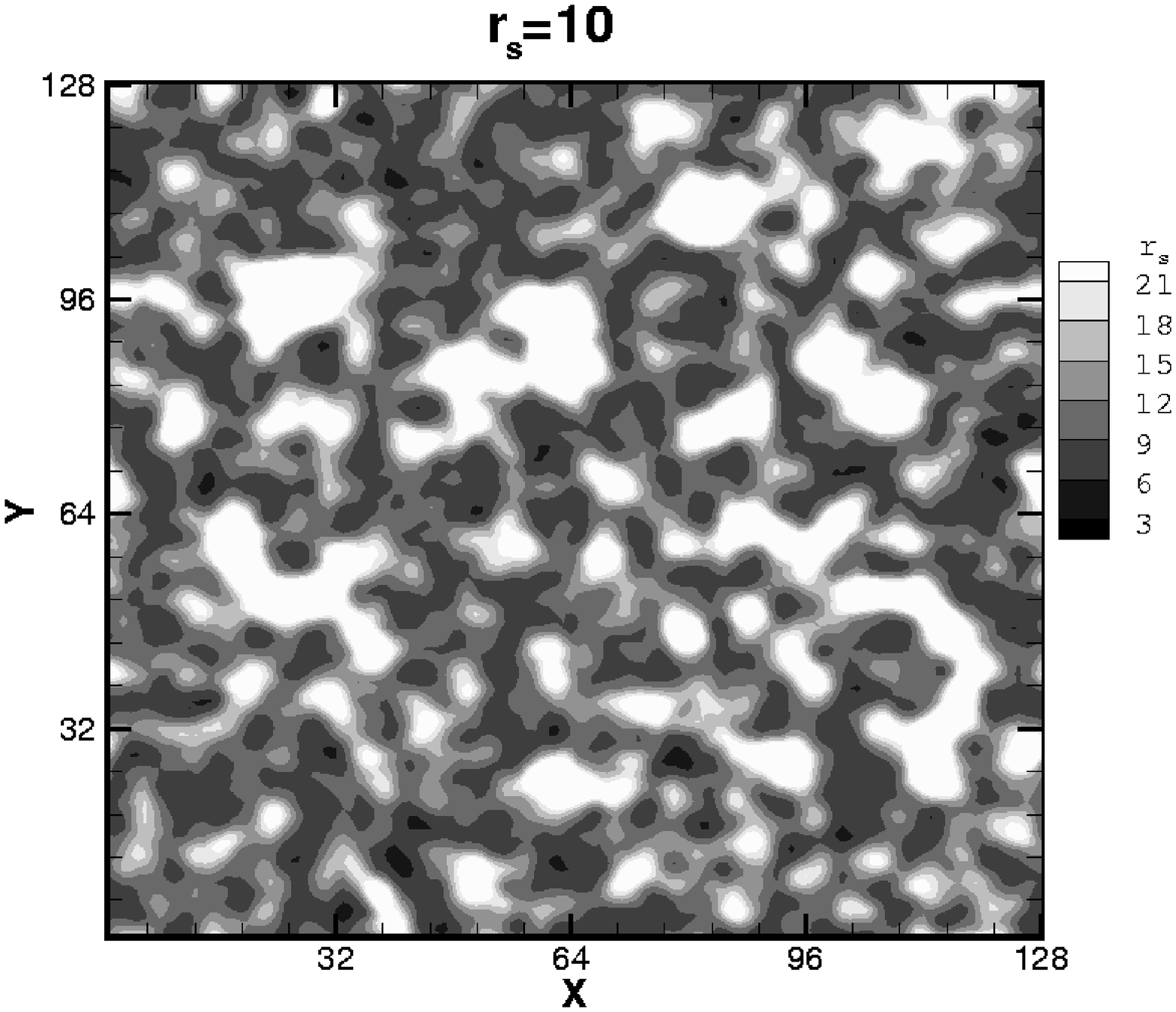, width=0.8\columnwidth}
  \epsfig{file=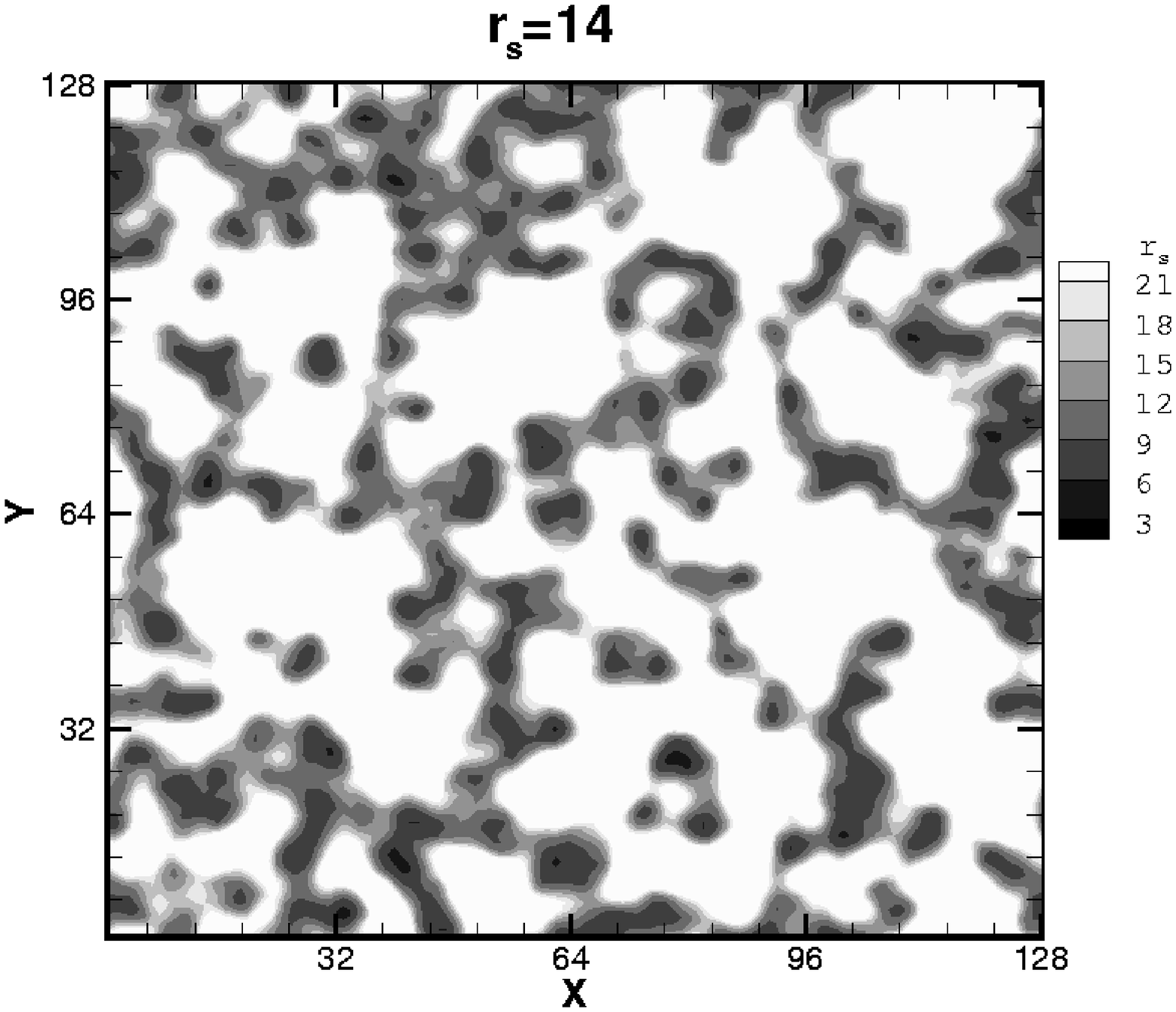, width=0.8\columnwidth}
  \epsfig{file=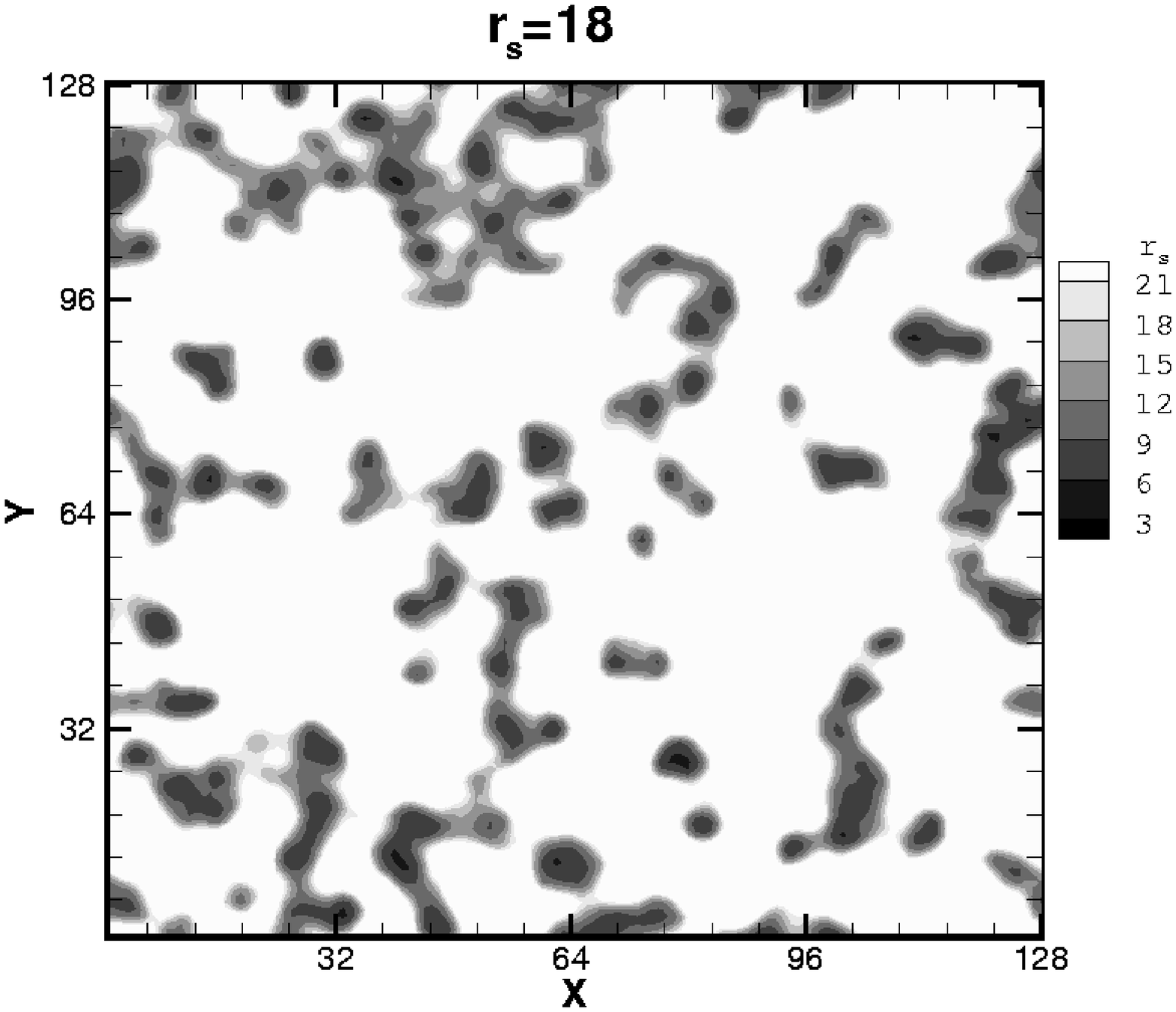, width=0.8\columnwidth}

  \caption{The density distributions for the different electron densities. 
    We use the contour plot for the local density parameter \protect\(
    r_{s}=1/\sqrt{\pi n}\protect \).  The density of the white area
    decreases rapidly to zero. We take \protect\( V_{s}=0.2Ry^*\protect
    \), \protect\( \xi =0.2L\protect \).}

  \label{distribution}
\end{figure}

Figure \ref{distribution} shows the density distribution of the system.
It can be clearly seen that the electrons form some high density
regions, while the density of other regions are essentially zero.
Depending on the average density of the system, the high density
regions may connect each other together ($ r_{s}=10 $), or form some
isolated regions ($ r_{s}=18 $). There exists a certain density ($
r_{s}=14$) where the high density regions starts to percolate through the
system, and form a conducting channel. The calculation clearly
demonstrates the idea of our earlier theory \cite{Shi}, \emph{i.e.,}
the metal-insulator transition observed in the 2D electron 
systems is the percolation transition of the electron
density.

\begin{figure}
  \centering
  \epsfig{file=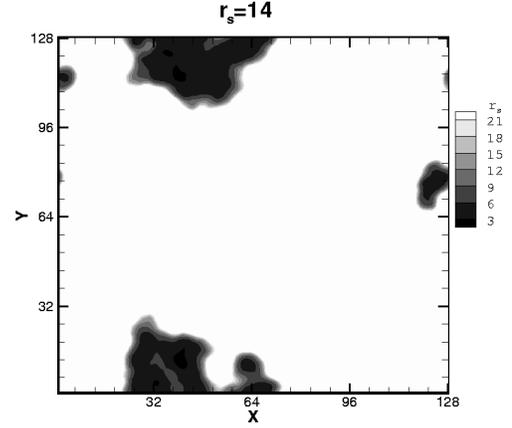, width=0.8\columnwidth}
  \caption{The density distribution for the free electron gas on the 
  same disorder landscape as the Fig.\ref{distribution} at density
  \protect\( r_{s}=14\protect \).}
\label{freedistribution}
\end{figure}

The electron-electron interaction is important for the conducting
behavior of a dilute electron system in the sense that it
makes the density distribution more extended because of
the Coulomb repulsion. 
Figure \ref{freedistribution} shows the density distribution for the free
electron gas with the same density as in Fig.1(b) by turning
off the electron-electron interaction. 
The system only forms some isolated high density regions
at the disorder valleys, while the density distribution of the
corresponding interacting system (Fig.1(b)) is quite extensive at the same
density.  At a given disorder strength, the critical density for the
free electron gas is much higher than its interacting counterpart.

\begin{figure}
  \centering
  \epsfig{file=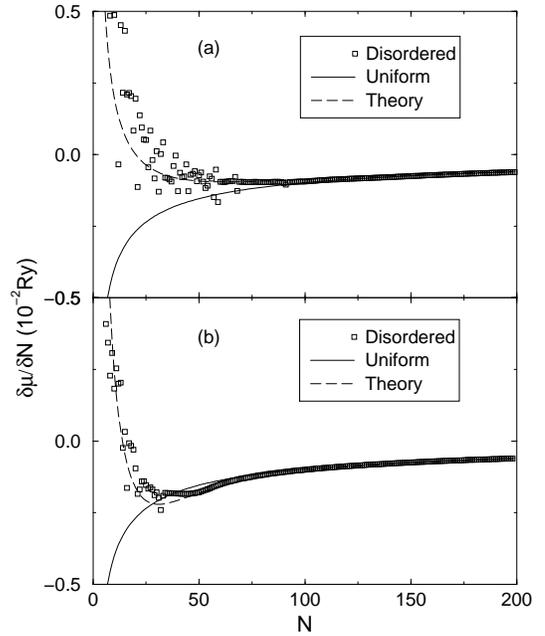, width=0.8\columnwidth}
  \caption{$ \delta \mu /\delta N\protect $ 
    as a function of the electron density. Solid lines are for
    the uniform electron gas, squares are the data points
    for the disordered system,
    and the dashed-lines are the results from the theory
    discussed in the text. Two kinds of
    the disorder are used in the calculation: (a) The same disorder
    landscape as in Figure \ref{distribution}; (b) Off-plane charge
    impurities with $d = 10a_B^*$, $n_{i}=2.5\times 10^{-3}/a_{B}^{*2}$. 
    The parameters in the dashed-lines:
    (a) $n_0 = 10^{-3}/a_{B}^{*2}$, $\alpha = 1.5$; 
    (b) $n_0 =0.5\times 10^{-3}/a_{B}^{*2}$, $\alpha = 2.3$.
    $N$ is the total number of
    the electrons in the simulation box.  $N=30$ corresponds to
    $r_{s}=18\protect$; $N=50$ corresponds to $r_{s}=14$; $N=100$
    corresponds to $r_{s}=10$.}
  \label{compressibility}
\end{figure}

Figure \ref{compressibility} shows the compressibility of the systems.
To compare with the experiments\cite{Jiang,Eisenstein}, 
we calculate $ \delta \mu
/\delta N $, which is the direct measured quantity in the experiments.
It is well known that the compressibility of a uniform electron
gas is negative in the low density region due to the effect of the
exchange and correlation energies, as shown by the solid line
in Fig.3(a). However,
when the disorder is present, the behavior changes greatly.  In the
low density, the electrons tend to occupy the valleys of the disorder
landscape, and the local density, instead of the average density,
determines the compressibility of the system. On the other hand,
at higher densities, all of the valleys are filled, one can expect the
compressibility of the system to resume the behavior of a uniform
electron gas. We have a non-monotonic behavior for $ \delta \mu
/\delta N $, as shown by the dots in Fig.3(a), which are in good
agreement with
the experimental measurement \cite{Jiang,Eisenstein}.  Comparing with 
Fig.\ref{distribution}, we find that the turning point of the
compressibility ($ N\approx 50 $, $ r_{s}\approx 14 $) coincides with
the percolation threshold of the system. At low densities, the data
points in the plot show strong fluctuation, indicating the effect of
the local fluctuation of the disorder potential.

The behavior can be understood by a simple theory. Following the
definition, the chemical potential $\mu$ is the energy needed to 
add an electron into the system, 
\begin{eqnarray*}
\mu(N) &=& E(N+1) - E(N) \\
       &\approx& \varepsilon_0\left[n_{eff}(N+1)\right](N+1) - 
           \varepsilon_0\left[n_{eff}(N)\right]N \\
       &=& \frac{\delta }{\delta N}
       \left\{ \varepsilon_0\left[n_{eff}(N)\right]N \right\},
\end{eqnarray*}
where we suppose that the electron energy is determined by the local
density of the electrons. $E(N)$ is the total energy of the system,
$\varepsilon_0(n)$ is the energy per electron for the uniform electron
gas, and $n_{eff}$ is effective local density. For the inhomogeneous
system as shown in Fig.\ref{distribution}, the effective local density
can be estimated by $n_{eff}(n) \approx n / f(n)$, where $f(n)$ is the
fraction of the high density region. After some algebra, we have
\begin{eqnarray*}
\mu(n) &=& \mu _{0}\left( \frac{n}{f}\right) 
\left[ 1-\frac{d\ln f}{d\ln n}\right] +\varepsilon _{0}
\left( \frac{n}{f}\right) \frac{d\ln f}{d\ln n},\\ 
\frac{\delta \mu}{\delta n} &=& \mu _{0}^\prime
\left( \frac{n}{f}\right) \frac{1}{f}
\left[ 1-\frac{d\ln f}{d\ln n}\right] ^{2}
-\varepsilon_{0}^\prime \left( \frac{n}{f}\right) 
\left( \frac{n}{f}\right) ^{2}\frac{d^{2}f}{dn^{2}} \, ,
\end{eqnarray*}
where $\mu_0$ is the chemical potential for a uniform electron gas.
In the low density limit, $f(n) \rightarrow 0$, the local effective
density is greatly different from the average density of the system.
As a consequence, the density dependence of the chemical potential,
$\delta \mu / \delta n$, changes greatly. In general,
supposing $f(n) \sim n^\alpha$ in the low density limit, the analysis
shows that $\delta \mu / \delta n$ will have a non-monotonic behavior if
$\alpha > 1$. The behavior of $f(n)$ is determined
by the local disorder potential profile. In a 2D system, the infinite harmonic
potential has $f(n) \sim n$. So the requirement $\alpha >1$ is
equivalent to the condition that the local disorder potential has a
weaker confinement effect than the harmonic potential. The condition
can be easily satisfied in a typical experimental system. For
instance, the coulomb potential has $f(n) \sim n^4$.  In 
Fig.\ref{compressibility}, we use the above equation 
for $\delta \mu / \delta n$
with the following relation of $f(n)$ to fit the data,
$$
f(n) = \frac{1}{1+\left(\frac{n_0}{n}\right)^\alpha}\, .
$$
This form has a correct behavior in the 
high density limit, $f \rightarrow 1$, and the low
density behavior is controlled by $\alpha$. By carefully choosing the
values for $\alpha$ and $n_0$,
we obtain a good agreement with our
numerical data as shown in the dashed line in Fig.3(a).

To demonstrate the effect of a different disorder potential, we use
off-plane charge impurity potential in calculating Figure
\ref{compressibility}(b), {\it i.e.},
$$
V_d({\bf x}) = -\sum_i \frac{1}{\sqrt{|{\bf x}-{\bf x}_i|^2 + d^2} } \, , 
$$
where $d$ is the distance between the electron and the impurity
planes, and the impurities are randomly distributed with
a density $n_{i}$. This potential gives a similar density
profile as shown in Figure 1. 
As expected, this form of potential has a larger value
of $\alpha$ as given in the figure caption.

To conclude, we have studied the electron space distribution and the
compressibility of disordered dilute 2D electron systems by using the
local density approximation. Electron distribution confirms
the formation of the droplet state that consists of high
and low density regions. Our calculated compressibility 
is in good agreement with the experimentally observed 
behavior showing unexpected anomaly at low densities.
The turning point of the compressibility happens around
the percolation threshold.
Our theory based on the droplet state provides a good understanding
of the compressibility anomaly.

\bigskip

This work is supported by DOE.

\end{document}